\documentstyle[12pt]{article}

\begin{document}

\title{\bf Electromagnetic Interaction Equations}

\author{Yury M. Zinoviev\thanks{This work was supported in part by the Russian Foundation
 for Basic Research (Grant No. 07 - 01 - 00144) and the Program for Supporting Leading
 Scientific Schools (Grant No. 795.2008.1).}}

\date{}
\maketitle

Steklov Mathematical Institute, Gubkin St. 8, 119991, Moscow,
Russia,

e - mail: zinoviev@mi.ras.ru

\vskip 1cm

\noindent {\bf Abstract.} For the electromagnetic interaction of two
particles the relativistic quantum mechanics equations are proposed.
These equations are solved for the case when one particle has a
small mass and moves freely. The initial wave functions are supposed
to be concentrated at the coordinates origin. The energy spectrum of
another particle wave function is defined by the initial wave
function of the free moving particle. Choosing the initial wave
functions of the free moving particle it is possible to obtain a
practically arbitrary energy spectrum.

\vskip 1cm

\section{Introduction}
\setcounter{equation}{0}

For the particles of the mass $\mu > 0$ and the spin $1/2$ the
relativistic equation was proposed by Dirac (\cite{2}, equation
(4.29); \cite{6}, equation (1 - 41); \cite{1}, equation (2.9))
\begin{equation}
\label{1.1} \sum_{n\, =\, 1}^{4} \sum_{\nu \, =\, 0}^{3}
(\gamma^{\nu})_{mn} \left( - i\frac{\partial}{\partial x^{\nu}}
\right) \psi_{n} (x) + \mu \psi_{m} (x) = 0.
\end{equation}
Let us consider a classical electromagnetic field with a vector
potential $A_{\nu}(x)$. If we change in the equation (\ref{1.1}) the
differential operator $- i\partial_{\nu}$ for the differential
operator $- i\partial_{\nu} + qA_{\nu}(x)$, we get the equation for
the interaction between the particle with the charge $q$ and the
external classical electromagnetic field (\cite{2}, Chapter 4,
equation (4.202); \cite{1}, Chapter 2, equation (2.62))
\begin{equation}
\label{1.2} \sum_{n\, =\, 1}^{4} \sum_{\nu \, =\, 0}^{3}
(\gamma^{\nu})_{mn} \left( - i\frac{\partial}{\partial x^{\nu}} +
qA_{\nu}(x) \right) \psi_{n} (x) + \mu \psi_{m} (x) = 0.
\end{equation}
The substitution of the electron charge $q = - e$ and the Coulomb
vector potential $A_{0}(x) = Ze(4\pi |{\bf x}|)^{- 1}$, $A_{k}(x) =
0$, $k = 1,2,3$, into the equation (\ref{1.2}) yields the equation
for the electron in the external electromagnetic field generated by
the nucleus with the charge $Ze$. We consider a nucleus as a
classical particle. The energy spectrum of this equation (\ref{1.2})
is discrete and the energy level $E_{1,1/2} = \mu c^{2}(1 -
(Z/137)^{2})^{1/2}$ (\cite{1}, equation (2.87)). For $Z
> 137$ the value $E_{1,1/2}$ becomes imaginary. The description of the
electromagnetic interaction between the electron and the nucleus
with the charge $Ze > 137e$ seems not to make sense. At present time
the nuclei with the charges $Ze \leq 118e$ are synthesized. (The
nucleus with the charge $Ze = 117e$ has not been yet discovered.)
The equation (\ref{1.2}) should be mathematically self - consistent
for Coulomb vector potential with any charge. Due to the paper
\cite{4} Hamiltonian (\ref{1.2}) with the Coulomb vector potential
$A_{0}(x) = Ze(4\pi |{\bf x}|)^{- 1}$, $A_{k}(x) = 0$, $k = 1,2,3$,
has a self adjoint operator extension for any charge $Ze$. For $Ze
\geq (\sqrt 3/2)137e
> 118e$ this self adjoint operator extension is not unique.

The equation (\ref{1.2}) defines the wave function of the electron
interacting with the electromagnetic field generated by the
classical particle. Both interacting particles should be quantum. In
the quantum electrodynamics (\cite{8}, Lecture 24) the following
equations for the electromagnetic interaction of two particles are
studied
\begin{eqnarray}
\label{1.3} \sum_{n_{1},n_{2}\, =\, 1}^{4} \Biggl( \left( \prod_{s\,
=\, 1}^{2} \left( \sum_{\nu \, =\, 0}^{3}
(\gamma^{\nu})_{m_{s},n_{s}} \left( i\frac{\partial}{\partial
x_{s}^{\nu}}\right) - \mu_{s} \delta_{m_{s},n_{s}} \right) \right)
\psi_{n_{1}n_{2},p_{1}p_{2}}
(x_{1},x_{2}) +  \sum_{\nu_{1}, \nu_{2} \, =\, 0}^{3} \nonumber \\
\eta_{\nu_{1} \nu_{2}} Kq_{1}q_{2}D_{0}^{c}(x_{1} - x_{2})\left(
\prod_{s\, =\, 1}^{2} (\gamma^{\nu_{s}})_{m_{s},n_{s}}\right)
\psi_{n_{1}n_{2},p_{1}p_{2}} (x_{1},x_{2})\Biggr) = i\prod_{s\, =\,
1}^{2} \delta (x_{s}) \delta_{m_{s},p_{s}},
\end{eqnarray}
\begin{eqnarray}
\label{1.4} D_{m^{2}}^{c}(x) = \lim_{\epsilon \, \rightarrow \, + 0}
(2\pi)^{- 4} \int d^{4}k \exp \{ i(k,x)\} (m^{2} -
(k,k) - i\epsilon )^{- 1}, \nonumber \\
(x,y) = x^{0}y^{0} - \sum_{k\, =\, 1}^{3} x^{k}y^{k},
\end{eqnarray}
where $q_{2}$, $q_{2}$ are the charges, $K$ is the electromagnetic
interaction constant, $\eta_{\mu \nu} = \eta^{\mu \nu}$ is the
diagonal matrix with diagonal matrix elements $\eta_{00} = -
\eta_{11} = - \eta_{22} = - \eta_{33} = 1$.

The solutions of the equations (\ref{1.3}) contain the divergent
integrals \cite{8}. The finite answers for the divergent integrals
are obtained by means of the renormalization procedure. Due to the
book (\cite{9}, Chapter 4): "The shell game that we play to find $n$
and $j$ is technically called "renormalization". But no matter how
clever the word, it is what I would call a dippy process! Having to
resort to such hocus - pocus has prevented us from proving that the
theory of quantum electrodynamics is mathematically self -
consistent. It's surprising that the theory still hasn't been proved
self - consistent one way or the other by now; I suspect that
renormalization is not mathematically legitimate. What {\it is}
certain is that we do not have a good mathematical way to describe
the theory of quantum electrodynamics: such a bunch of words to
describe the connection between $n$ and $j$ and $m$ and $e$ is not
good mathematics."

The equations (\ref{1.3}) do not satisfy the causality condition:
the support of the fundamental solution $D_{0}^{c}(x)$ of the wave
equation does not lie in the upper or lower light cones. The
distribution $D_{0}^{c}(x)$ choice is probably connected with the
causality condition. The chronological product of two free scalar
field operators is defined as
\begin{equation}
\label{1.5} T(\phi(x) \phi(y)) = \phi(x) \phi(y) + ie_{m^{2}}(y -
x)I
\end{equation}
where $I$ is the identity operator and the distribution
\begin{equation}
\label{1.6} - e_{m^{2}}(- x) = \lim_{\epsilon \, \rightarrow \, +\,
0} (2\pi)^{- 4} \int d^{4}k \exp \{ i(k,x)\} (m^{2} - (k^{0} +
i\epsilon)^{2} + |{\bf k}|^{2})^{- 1}.
\end{equation}
The vacuum expectation of the product of two free scalar fields is
defined as
\begin{eqnarray}
\label{1.7} <\phi(x) \phi(y)>_{0} = - iD_{m^{2}}^{-}(x - y),
\nonumber \\ D_{m^{2}}^{-} (x) = i(2\pi)^{- 3} \int d^{4}k \exp \{ -
i(k,x)\} \theta (k^{0})\delta ((k,k) - m^{2}).
\end{eqnarray}
The vacuum expectation of the chronological product (\ref{1.5}) is
equal to
\begin{equation}
\label{1.8} <T(\phi(x) \phi(y))>_{0} = - iD_{m^{2}}^{-}(x - y) +
ie_{m^{2}}(y - x) = - iD_{m^{2}}^{c}(x - y).
\end{equation}
Stueckelberg and Rivier \cite{11} believed that the classical
"causal action" is given by the distribution $- e_{m^{2}}(y - x)$
and the distribution $D_{m^{2}}^{c}(x - y) = D_{m^{2}}^{c}(y - x)$
defines the probability amplitude of the "causal action".

The equations (\ref{1.3}) do not look like the equations
(\ref{1.2}). We want to write down two equations of the type
(\ref{1.2}) where one particle interacts with the electromagnetic
field generated by another particle. These equations should satisfy
the causality condition. Let us define $4\times 4$ - matrices
\begin{equation}
\label{3.17} \alpha (\mu^{2}) = \left( \begin{array}{cc}

\mu^{2} \sigma^{0} & 0 \\

0 & \sigma^{0}

\end{array} \right),  \, \,
\beta (\mu^{2}) = \left( \begin{array}{cc}

\sigma^{0} & 0 \\

0 & \mu^{2} \sigma^{0}

\end{array} \right), \, \,
\gamma^{\nu} = \left( \begin{array}{cc}

0 & \eta^{\nu \nu} \sigma^{\nu} \\

\sigma^{\nu} & 0

\end{array} \right), \, \, \nu = 0,...,3,
\end{equation}
where $2\times 2$ - matrices $\sigma^{\nu}$ are given by the
relations
\begin{equation}
\label{11.3} \sigma^{0} = \left( \begin{array}{cc}

1 & 0 \\

0 & 1

\end{array} \right),
\sigma^{1} = \left( \begin{array}{cc}

0 & 1 \\

1 & 0

\end{array} \right),
\sigma^{2} = \left( \begin{array}{cc}

0 & - i \\

i &   0

\end{array} \right),
\sigma^{3} = \left( \begin{array}{cc}

1 & 0 \\

0 & - 1

\end{array} \right).
\end{equation}
Let us consider two equations for the wave functions
$(\psi_{s})_{n_{s}} (x_{s})$, $s = 1,2$, equal to zero for the
negative $x^{0}_{s}$
\begin{eqnarray}
\label{1.10}  \sum_{n_{s}\, =\, 1}^{4} \left(  \sum_{\nu_{s} \, =\,
0}^{3} (\gamma^{\nu_{s}})_{m_{s}n_{s}} \left( -
i\frac{\partial}{\partial x_{s}^{\nu_{s}}} \right) + (\beta
(\mu_{s}^{2}))_{m_{s}n_{s}} \right) (\psi_{s})_{n_{s}} (x_{s}) =
\nonumber \\ - i\sum_{n_{s}\, =\, 1}^{4} (\gamma^{0})_{m_{s},n_{s}}
\delta (x_{s}^{0})(\psi_{s})_{n_{s}} (+ 0,{\bf x}_{s}),\, \, {\bf
x}_{s} = (x_{s}^{1},x_{s}^{2},x_{s}^{3}) \in {\bf R}^{3}, \, \, s =
1,2,
\end{eqnarray}
where $(\psi_{s})_{n_{s}} (+ 0,{\bf x}_{s})$ are the initial wave
functions. For $\mu_{1}, \mu_{2} > 0$ the equations (\ref{1.10})
with the right - hand sides equal to zero are equivalent to Dirac
equation (\ref{1.1}). For $\mu_{1}, \mu_{2} \geq 0$ the equations
(\ref{1.10}) with the right - hand sides equal to zero are
equivalent to the system of Klein - Gordon equations. The right -
hand sides of the equations (\ref{1.10}) are the standard method to
define Cauchy problem. The equations (\ref{1.10}) describe the free
motion of particles with the masses $\mu_{1}, \mu_{2} \geq 0$. In
contrast with the equation (\ref{1.1}) the particles with zero mass
are not distinguished. If the initial wave functions
$(\psi_{s})_{n_{s}} (+ 0,{\bf x}_{s}) = (\psi_{s})_{n_{s}} \delta
({\bf x}_{s})$, $s = 1,2$, then the product of the equations
(\ref{1.10}) looks likes the equation (\ref{1.3}) with the
interaction constant $K = 0$.

The equation of the electromagnetic action of the second particle on
the first particle has the form
\begin{eqnarray}
\label{1.9} \sum_{n_{s} \, =\, 1,...,4,\atop s\, =\, 1,2}  \int
d^{4}x_{2} \Biggl( \prod_{s\, =\, 1}^{2} \left( \sum_{\nu \, = \,
0}^{3} (\gamma^{\nu})_{m_{s}n_{s}} \left( -
i\frac{\partial}{\partial x_{s}^{\nu}}\right) + (\beta
(\mu_{s}^{2}))_{m_{s}n_{s}} \right) (\psi_{s})_{n_{s}} (x_{s}) +
\nonumber \\ \sum_{\nu_{1}, \nu_{2} \, =\, 0}^{3}
Kq_{1}q_{2}\eta_{\nu_{1} \nu_{2}} e_{0}(x_{1} - x_{2}) \prod_{s\,
=\, 1}^{2} (\gamma^{\nu_{s}}
)_{m_{s}n_{s}} (\psi_{s})_{n_{s}} (x_{s})\Biggr) = \nonumber \\
- \sum_{n_{s} \, =\, 1,...,4,\atop s\, =\, 1,2}  \int d^{4}x_{2}
\prod_{s\, =\, 1}^{2} (\gamma^{0})_{m_{s} n_{s}} \delta
(x_{s}^{0})(\psi_{s})_{n_{s}} (+ 0,{\bf x}_{s}).
\end{eqnarray}
The equation of the electromagnetic action of the first particle on
the second particle is the similar one. Both equations of the type
(\ref{1.2}) satisfy the causality condition. The support of the
distribution $- e_{0}(- x)$ lies in the closed lower light cone. The
fundamental solution $- e_{0}(- x)$ of the wave equation is unique
in the class of the distributions with supports in the lower light
cone. The causality condition defines the distribution $- e_{0}(-
x)$ uniquely. The distribution (\ref{1.6}) differs from the
distribution (\ref{1.4}) in the rule of going around the poles in
the integral. The support of the distribution $- e_{0}(- x)$ lies on
the lower light cone boundary.  The distribution $- e_{0}(- x)$
gives the delay. It is necessary to have the delay in a relativistic
interaction equation. It was already noted by Poincar\'e \cite{10}.
The interaction propagates not instantly but at the speed of light.
We have to take into account the distance between the interacting
particles. The delay is one of possible causality statements.

In this paper the equations (\ref{1.9}) are solved for the case when
the second particle has a small mass and the wave function
$(\psi_{2})_{n_{2}} (x_{2})$ satisfies the second equation
(\ref{1.10}) for the initial wave function $(\psi_{2})_{n_{2}} (+
0,{\bf x}_{2}) = (\psi_{2})_{n_{2}} \delta ({\bf x}_{2})$. The
energy spectrum of the solutions $(\psi_{1})_{n_{1}} (x_{1})$ is
defined by the initial wave function $(\psi_{2})_{n_{2}} \delta
({\bf x}_{2})$. By making a choice of the vectors
$(\psi_{2})_{n_{2}}$ it is possible to obtain a practically
arbitrary energy spectrum.

\section{Dirac equation}
\setcounter{equation}{0}

It is necessary to write down an equation of the type (\ref{1.1})
for the free motion of a particle with arbitrary spin and mass. Let
us consider the complex $2\times 2$ - matrices
\begin{equation}
\label{11.1} A =  \left( \begin{array}{cc}

A_{11} & A_{12} \\

A_{21} & A_{22}

\end{array} \right).
\end{equation}
The $2\times 2$ - matrix
\begin{equation}
\label{11.2} A^{\ast} = \left( \begin{array}{cc}

\bar{A}_{11} & \bar{A}_{21} \\

\bar{A}_{12} & \bar{A}_{22}

\end{array} \right)
\end{equation}
is called Hermitian conjugate. If $A^{\ast} = A$, the matrix
(\ref{11.1}) is Hermitian. The matrices (\ref{11.3}) form a basis of
Hermitian matrices. The multiplication rules for the matrices
(\ref{11.3}) are
\begin{eqnarray}
\label{11.5} \sigma^{\mu} \sigma^{\mu} = \sigma^{0},\, \, \sigma^{0}
\sigma^{\mu} = \sigma^{\mu} \sigma^{0} = \sigma^{\mu}, \, \, \mu =
0,...,3; \nonumber \\
\sigma^{k_{1}} \sigma^{k_{2}} = \sum_{k_{3}\, =\, 1}^{3}
\epsilon^{k_{1}k_{2}k_{3}} i \sigma^{k_{3}}, \, \, k_{1},k_{2} =
1,2,3, \, \, k_{1} \neq k_{2}
\end{eqnarray}
where the antisymmetric tensor $\epsilon^{k_{1}k_{2}k_{3}}$ has the
normalization $\epsilon^{123} = 1$.

We identify the four dimensional Minkowski space with the four
dimensional space of Hermitian $2\times 2$ - matrices
\begin{equation}
\label{11.10} \tilde{x} = \sum_{\mu = 0}^{3} x^{\mu} \sigma^{\mu}.
\end{equation}
For a complex $2\times 2$ - matrix (\ref{11.1}) we define the
following $2\times 2$ - matrices
\begin{equation}
\label{2.1} A^{T} = \left( \begin{array}{cc}

A_{11} & A_{21} \\

A_{12} & A_{22}

\end{array} \right), \, \,
\bar{A} = \left( \begin{array}{cc}

\bar{A}_{11} & \bar{A}_{12} \\

\bar{A}_{21} & \bar{A}_{22}

\end{array} \right).
\end{equation}
The matrices (\ref{11.1}) with determinant equal to $1$ form the
group $SL(2,{\bf C})$. The matrices (\ref{11.1}) satisfying the
equations $A^{\ast}A = \sigma^{0}$, $\det A = 1$ form the group
$SU(2)$. The group $SU(2)$ is the maximal compact subgroup of the
group $SL(2,{\bf C})$. Let us describe the irreducible
representations of the group $SU(2)$. We consider the non - negative
half - integers $l = 0,1/2,1,3/2,...$. We define the representation
of the group $SU(2)$ on the space of the polynomials with degrees
less than or equal to $2l$
\begin{equation}
\label{2.2} T_{l}(A)\phi (z) = (A_{12}z + A_{22})^{2l}\phi \left(
 \frac{A_{11}z + A_{21}}{A_{12}z + A_{22}}\right).
\end{equation}
We consider a half - integer $n = - l, - l + 1,...,l - 1,l$ and
choose the polynomial basis
\begin{equation}
\label{2.3} \psi_{n} (z) = ((l - n)!(l + n)!)^{- 1/2}z^{l - n}.
\end{equation}
The definitions (\ref{2.2}), (\ref{2.3}) imply
\begin{equation}
\label{2.4} T_{l}(A)\psi_{n} (z) = \sum_{m = - l}^{l} \psi_{m}
(z)t_{mn}^{l}(A),
\end{equation}
\begin{eqnarray}
\label{2.5} t_{mn}^{l}(A) = ((l - m)!(l + m)!(l - n)!(l + n)!)^{1/2}
\times \nonumber \\ \sum_{j = - \infty}^{\infty} \frac{A_{11}^{l - m
- j}A_{12}^{j}A_{21}^{m - n + j}A_{22}^{l + n - j}}{\Gamma (j +
1)\Gamma (l - m - j + 1)\Gamma (m - n + j + 1)\Gamma (l + n - j +
1)}
\end{eqnarray}
where $\Gamma (z)$ is the gamma - function. The function $(\Gamma
(z))^{- 1}$ equals zero for $z = 0 , - 1, - 2,...$. Therefore the
series (\ref{2.5}) is the polynomial.

The relation (\ref{2.2}) defines the representation of the group
$SU(2)$. Thus the polynomial (\ref{2.5}) defines the representation
of the group $SU(2)$
\begin{equation}
\label{2.6} t_{mn}^{l}(AB) = \sum_{k = - l}^{l}
t_{mk}^{l}(A)t_{kn}^{l}(B).
\end{equation}
This $(2l + 1)$ - dimensional representation is irreducible
(\cite{5}, Chapter III, Section 2.3). The relations (\ref{2.5}),
(\ref{2.6}) have an analytic continuation to all matrices
(\ref{11.1}).

By making the change $j \rightarrow j + n - m$ of the summation
variable in the equality (\ref{2.5}) we have
\begin{equation}
\label{2.7} t_{mn}^{l}(A) = t_{nm}^{l}(A^{T}).
\end{equation}
The definition (\ref{2.5}) implies
\begin{equation}
\label{2.10} t_{mn}^{l}(\sigma^{0}) = \delta_{mn},
\end{equation}
The polynomial (\ref{2.5}) is homogeneous of the matrix elements
(\ref{11.1}). Its degree is $2l$. The sum (\ref{11.5}) contains the
only non - zero term. The relations (\ref{11.5}), (\ref{2.6}) imply
\begin{eqnarray}
\label{2.8} \sum_{p\, =\, - l}^{l} \sum_{\dot{p} \, =\, -
\dot{l}}^{\dot{l}} \left( \sum_{\nu \, =\, 0}^{3}
t_{mp}^{l}(\sigma^{\nu})t_{\dot{m}
\dot{p}}^{\dot{l}}(\overline{\sigma^{\nu}}) \left( -
i\frac{\partial}{\partial x^{\nu}}\right) \right) \left( \sum_{\nu
\, =\, 0}^{3} \eta^{\nu \nu} t_{pn}^{l}(\sigma^{\nu})t_{\dot{p}
\dot{n}}^{\dot{l}}(\overline{\sigma^{\nu}})
\left( - i\frac{\partial}{\partial x^{\nu}} \right) \right) = \nonumber \\
\sum_{{1\, \leq \, k_{1}\, <\, k_{2}\, \leq \, 3,} \atop k_{3}\, =\,
1,...,3}  t_{mn}^{l}(i\sigma^{k_{3}})t_{\dot{m} \dot{n}}^{\dot{l}}(-
i\overline{\sigma^{k_{3}}})((\epsilon^{k_{1}k_{2}k_{3}})^{2l +
2\dot{l}} + (\epsilon^{k_{2}k_{1}k_{3}})^{2l +
2\dot{l}})\frac{\partial^{2}}{\partial x^{k_{1}}\partial x^{k_{2}}}
\nonumber \\ - \delta_{mn} \delta_{\dot{m} \dot{n}} (\partial_{x},
\partial_{x}),
\end{eqnarray}
\begin{equation}
\label{2.810} (\partial_{x},
\partial_{x}) = \sum_{\nu \, =\, 0}^{3} \eta^{\nu \nu} \left(
\frac{\partial}{\partial x^{\nu}} \right)^{2}.
\end{equation}
For an odd integer $2l + 2\dot{l}$ the relation (\ref{2.8}) has the
form
\begin{eqnarray}
\label{2.9} \sum_{p\, =\, - l}^{l} \sum_{\dot{p} \, =\, -
\dot{l}}^{\dot{l}} \left( \sum_{\nu \, =\, 0}^{3}
t_{mp}^{l}(\sigma^{\nu})t_{\dot{m}
\dot{p}}^{\dot{l}}(\overline{\sigma^{\nu}}) \left( -
i\frac{\partial}{\partial x^{\nu}}\right) \right) \left( \sum_{\nu
\, =\, 0}^{3} \eta^{\nu \nu} t_{pn}^{l}(\sigma^{\nu})t_{\dot{p}
\dot{n}}^{\dot{l}}(\overline{\sigma^{\nu}})\left( -
i\frac{\partial}{\partial x^{\nu}}\right) \right) = \nonumber \\  -
\delta_{mn} \delta_{\dot{m} \dot{n}} (\partial_{x},
\partial_{x}).
\end{eqnarray}
By making use of the relation (\ref{2.9}) for an odd integer $2l +
2\dot{l}$ Lorentz invariant equations
\begin{equation}
\label{2.11} ( - (\partial_{x}, \partial_{x}) - \mu^{2})
\phi_{m\dot{m}} (x) = 0,
\end{equation}
$m = - l, - l + 1,...,l - 1,l$, $\dot{m} = - \dot{l}, - \dot{l} +
1,...,\dot{l} - 1,\dot{l}$, may be rewritten as the system of the
linear equations
\begin{eqnarray}
\label{2.12} \sum_{n\, =\, 2l + 2}^{4l + 2} \sum_{\dot{n} \, =\,
1}^{2\dot{l} + 1} \sum_{\nu\, =\, 0}^{3} \eta^{\nu \nu} t_{m - l -
1,n - 3l - 2}^{l}(\sigma^{\nu}) t_{\dot{m} - \dot{l} - 1,\dot{n} -
\dot{l} - 1}^{\dot{l}}(\overline{\sigma^{\nu}})\left( -
i\frac{\partial}{\partial x^{\nu}} \right) \psi_{n\dot{n}} (x) +
\nonumber \\ \psi_{m\dot{m}} (x) = 0,\, \,  m = 1,..., 2l + 1,\, \,
\dot{m} = 1,...,2\dot{l} + 1;\nonumber \\ \sum_{n\, =\, 1}^{2l + 1}
\sum_{\dot{n} \, =\, 1}^{2\dot{l} + 1} \sum_{\nu\, =\, 0}^{3} t_{m -
3l - 2,n - l - 1}^{l}(\sigma^{\nu}) t_{\dot{m} - \dot{l} - 1,\dot{n}
- \dot{l} - 1}^{\dot{l}}(\overline{\sigma^{\nu}})\left( -
i\frac{\partial}{\partial x^{\nu}} \right) \psi_{n\dot{n}} (x) +
\nonumber \\ \mu^{2} \psi_{m\dot{m}} (x) = 0,\, \,  m = 2l + 2,...,
4l + 2,\, \, \dot{m} = 1,...,2\dot{l} + 1.
\end{eqnarray}
Let us define the $((4l + 2)(2\dot{l} + 1))\times ((4l + 2)(2\dot{l}
+ 1))$ - matrices
\begin{eqnarray}
\label{2.13} (\alpha_{l,\dot{l}} (\mu^{2}))_{m\dot{m},n\dot{n}} =
\mu^{2} \delta_{mn} \delta_{\dot{m} \dot{n}}, \, \, m,n = 1,...,2l +
1,\, \, \dot{m}, \dot{n} = 1,...,2\dot{l} + 1;\nonumber
\\ (\alpha_{l,\dot{l}} (\mu^{2}))_{m\dot{m},n\dot{n}} =
\delta_{mn} \delta_{\dot{m} \dot{n}}, \, \, m,n = 2l + 2,...,4l +
2,\, \, \dot{m}, \dot{n} = 1,...,2\dot{l} + 1; \nonumber \\
(\beta_{l,\dot{l}} (\mu^{2}))_{m\dot{m},n\dot{n}} = \delta_{mn}
\delta_{\dot{m} \dot{n}}, \, \, m,n = 1,...,2l +
1,\, \, \dot{m}, \dot{n} = 1,...,2\dot{l} + 1;\nonumber \\
(\beta_{l,\dot{l}} (\mu^{2}))_{m\dot{m},n\dot{n}} = \mu^{2}
\delta_{mn} \delta_{\dot{m} \dot{n}}, \, \, m,n = 2l + 2,...,4l +
2,\, \, \dot{m}, \dot{n} = 1,...,2\dot{l} + 1; \nonumber
\\ (\gamma_{l,\dot{l}}^{\nu} (\sigma^{0}))_{m\dot{m}, n\dot{n}} =
\eta^{\nu \nu} t_{m - l - 1,n - 3l - 2}^{l}(\sigma^{\nu}) t_{\dot{m}
- \dot{l} - 1,\dot{n} - \dot{l} -
1}^{\dot{l}}(\overline{\sigma^{\nu}}),\nonumber \\ m = 1,...,2l +
1,\, \, n = 2l + 2,...,4l + 2,\, \, \dot{m}, \dot{n} = 1,...,
2\dot{l} + 1,\, \, \nu = 0,...,3; \nonumber
\\ (\gamma_{l,\dot{l}}^{\nu} (\sigma^{0}))_{m\dot{m}, n\dot{n}} =
t_{m - 3l - 2,n - l - 1}^{l}(\sigma^{\nu}) t_{\dot{m} - \dot{l} -
1,\dot{n} - \dot{l} -
1}^{\dot{l}}(\overline{\sigma^{\nu}}),\nonumber
\\ m = 2l + 2,...,4l + 2,\, \, n = 1,...,2l + 1,\, \, \dot{m},
\dot{n} = 1,..., 2\dot{l} + 1,\, \, \nu = 0,...,3.
\end{eqnarray}
The other matrix elements are equal to zero. If an integer $2l +
2\dot{l}$ is odd, then an integer $(2l + 1)(2\dot{l} + 1) =
4l\dot{l} + 2l + 2\dot{l} + 1$ is even and $4$ divides into $(4l +
2)(2\dot{l} + 1)$.  For $l = 1/2$, $\dot{l} = 0$ the integer $(4l +
2)(2\dot{l} + 1) = 4$. The definitions (\ref{2.13}) are the
straightforward generalizations of the definitions (\ref{3.17}). The
definition (\ref{2.5}) implies
\begin{equation}
\label{2.17} t_{mn}^{\frac{1}{2}} (A) = A_{m + \frac{3}{2}, n +
\frac{3}{2}}.
\end{equation}
$4\times 4$ - matrices $\alpha_{1/2,0} (\mu^{2})$, $\beta_{1/2,0}
(\mu^{2})$, $\gamma_{1/2,0}^{\nu} (\sigma^{0})$ coincide with the
matrices $\alpha (\mu^{2})$, $\beta (\mu^{2})$, $\gamma^{\nu}$ given
by the relations (\ref{3.17}). By making use of the definitions
(\ref{2.13}) we can rewrite the equations (\ref{2.12}) as
\begin{equation}
\label{2.14} \sum_{n\, =\, 1}^{4l + 2} \sum_{\dot{n} \, =\,
1}^{2\dot{l} + 1} \left( \sum_{\nu \, =\, 0}^{3}
(\gamma_{l,\dot{l}}^{\nu} (\sigma^{0}))_{m\dot{m}, n\dot{n}} \left(
- i\frac{\partial}{\partial x^{\nu}} \right) + (\beta_{l,\dot{l}}
(\mu^{2}))_{m\dot{m}, n\dot{n}} \right) \psi_{n\dot{n}} (x) = 0.
\end{equation}
In contrast with the equation (\ref{1.1}) the mass $\mu$ in the
equation (\ref{2.14}) may be equal to zero. The definitions
(\ref{2.13}) imply
\begin{eqnarray}
\label{2.15} \alpha_{l,\dot{l}} (\mu)
\gamma_{l,\dot{l}}^{\nu} (\sigma^{0}) = \gamma_{l,\dot{l}}^{\nu}
(\sigma^{0}) \beta_{l,\dot{l}} (\mu), \, \,
\beta_{l,\dot{l}} (\mu^{2}) \gamma_{l,\dot{l}}^{\nu} (\sigma^{0}) =
\gamma_{l,\dot{l}}^{\nu} (\sigma^{0}) \alpha_{l,\dot{l}} (\mu^{2}),
\nonumber \\
\alpha_{l,\dot{l}} (\mu) \beta_{l,\dot{l}} (\mu^{2}) =
\mu  \beta_{l,\dot{l}} (\mu).
\end{eqnarray}
In view of the relations (\ref{2.15})  the action of the matrix
$\alpha_{l,\dot{l}} (\mu)$ on the equation (\ref{2.14}) yields the
equation of type (\ref{1.1})
\begin{equation}
\label{2.161} \sum_{n\, =\, 1}^{4l + 2} \sum_{\dot{n} \, =\,
1}^{2\dot{l} + 1}  \sum_{\nu \, =\, 0}^{3} (\gamma_{l,\dot{l}}^{\nu}
(\sigma^{0}))_{m\dot{m}, n\dot{n}} \left( -
i\frac{\partial}{\partial x^{\nu}} \right) \xi_{n\dot{n}} (x) + \mu
\xi_{m\dot{m}} (x) = 0,
\end{equation}
\begin{equation}
\label{2.162} \xi_{m\dot{m}} (x) = \sum_{n\, =\, 1}^{4l + 2}
\sum_{\dot{n} \, =\, 1}^{2\dot{l} + 1} (\beta_{l,\dot{l}}
(\mu))_{m\dot{m}, n\dot{n}} \psi_{n\dot{n}} (x) = 0.
\end{equation}
For $\mu > 0$ the transformation given by the relation (\ref{2.162})
is the isomorphism. Due to the relations (\ref{3.17}), (\ref{2.13}),
(\ref{2.17}) the equation (\ref{2.161}) for $l = 1/2$, $\dot{l} = 0$
coincides with Dirac equation (\cite{6}, equation (1 - 41)). The
relations (\ref{2.13}) imply
\begin{equation}
\label{2.401} (\alpha_{l,\dot{l}} (\mu^{2}) \beta_{l,\dot{l}}
(\mu^{2}))_{m\dot{m}, n\dot{n}} = (\beta_{l,\dot{l}}(\mu^{2})
\alpha_{l,\dot{l}} (\mu^{2}))_{m\dot{m}, n\dot{n}} = \mu^{2}
\delta_{mn} \delta_{\dot{m} \dot{n}},
\end{equation}
$m,n = 1,...,4l + 2,\, \dot{m}, \dot{n} = 1,...,2\dot{l} + 1$. In
view of the second relation (\ref{2.15}) and the relation
(\ref{2.401}) the action of the matrix $\gamma_{l,\dot{l}}^{0}
(\sigma^{0}) \alpha_{l,\dot{l}} (\mu^{2})$ on the equation
(\ref{2.14}) yields
\begin{equation}
\label{2.402} \sum_{n\, =\, 1}^{4l + 2} \sum_{\dot{n} \, =\,
1}^{2\dot{l} + 1} \left( \sum_{\nu \, =\, 0}^{3}
(\gamma_{l,\dot{l}}^{0} (\sigma^{0}) \gamma_{l,\dot{l}}^{\nu}
(\sigma^{0}) \beta_{l,\dot{l}} (\mu^{2}) )_{m\dot{m}, n\dot{n}}
\left( - i\frac{\partial}{\partial x^{\nu}}\right)  +  \mu^{2}
(\gamma_{l,\dot{l}}^{0} (\sigma^{0}) )_{m\dot{m}, n\dot{n}} \right)
\psi_{n\dot{n}} (x) = 0.
\end{equation}
The relations (\ref{11.5}), (\ref{2.6}), (\ref{2.10}), (\ref{2.13})
imply
\begin{equation}
\label{2.403} ((\gamma_{l,\dot{l}}^{0} (\sigma^{0}))^{2})_{m\dot{m},
n\dot{n}} = \delta_{mn} \delta_{\dot{m} \dot{n}}, \, \, m,n =
1,...,4l + 2,\, \, \dot{m}, \dot{n} = 1,...,2\dot{l} + 1,
\end{equation}
\begin{eqnarray}
\label{2.4031} (\gamma_{l,\dot{l}}^{0} (\sigma^{0})
\gamma_{l,\dot{l}}^{k} (\sigma^{0}))_{m\dot{m}, n\dot{n}} = -
(\gamma_{l,\dot{l}}^{k} (\sigma^{0}) \gamma_{l,\dot{l}}^{0}
(\sigma^{0}))_{m\dot{m}, n\dot{n}} = \nonumber \\ t_{m - l - 1,n - l
- 1}^{l} (\sigma^{k}) t_{m - \dot{l} - 1,n - \dot{l} - 1}^{\dot{l}}
(\overline{\sigma^{k}}), \nonumber \\ m,n = 1,...,2l + 1,\, \,
\dot{m}, \dot{n} = 1,...,2\dot{l} + 1,\, \, k = 1,2,3; \nonumber
\\ (\gamma_{l,\dot{l}}^{0} (\sigma^{0}) \gamma_{l,\dot{l}}^{k}
(\sigma^{0}))_{m\dot{m}, n\dot{n}} = - (\gamma_{l,\dot{l}}^{k}
(\sigma^{0}) \gamma_{l,\dot{l}}^{0} (\sigma^{0}))_{m\dot{m},
n\dot{n}} = \nonumber \\ - t_{m - 3l - 2,n - 3l - 2}^{l}
(\sigma^{k}) t_{m - \dot{l} - 1,n - \dot{l} - 1}^{\dot{l}}
(\overline{\sigma^{k}}), \nonumber \\ m,n = 2l + 2,...,4l + 2,\, \,
\dot{m}, \dot{n} = 1,...,2\dot{l} + 1,\, \, k = 1,2,3.
\end{eqnarray}
The other matrix elements are equal to zero. The coefficients of the
polynomial (\ref{2.5}) are real. Hence in view of the relations
(\ref{2.7}), (\ref{2.13}), (\ref{2.403}), (\ref{2.4031}) the
matrices $\gamma_{l,\dot{l}}^{0} (\sigma^{0})$,

\noindent $\gamma_{l,\dot{l}}^{0} (\sigma^{0})
\gamma_{l,\dot{l}}^{\nu} (\sigma^{0}) \beta_{l,\dot{l}} (\mu^{2})$,
$\nu = 0,...,3$, are Hermitian. Due to the relation (\ref{2.403}) we
have

\noindent $(\gamma_{l,\dot{l}}^{0} (\sigma^{0}))^{2}
\beta_{l,\dot{l}} (\mu^{2}) = \beta_{l,\dot{l}} (\mu^{2})$. Now the
equation (\ref{2.402}) implies that the integral
\begin{equation}
\label{2.404} \int d^{3}{\bf x} \sum_{m\, =\, 1}^{4l + 2}
\sum_{\dot{m} \, =\, 1}^{2\dot{l} + 1} (\beta_{l,\dot{l}}
(\mu^{2}))_{m\dot{m}, m\dot{m}} |\psi_{m\dot{m}} (x)|^{2}
\end{equation}
is independent of the variable $x^{0}$ for $x^{0} > 0$. The
integrand (\ref{2.404}) is called the probability density of a
solution of the equation (\ref{2.14}). For $\mu > 0$ the
transformation given by the relation (\ref{2.162}) is the
isomorphism. The probability density (\ref{2.404}) expressed through
the function (\ref{2.162}) is independent of the mass and coincides
with the usual probability density for the function (\ref{2.162}).
In the quantum mechanics the fixed probability density defines
Hilbert space where any Hamiltonian acts. The integral (\ref{2.404})
depends on the parameter $\mu^{2}$ in the equation (\ref{2.14}). We
do not want to suppose that all particles have strictly positive
masses. The experiments deal with the asymptotic solutions of the
interaction equations. We expect the solutions of the interaction
equations coincide asymptotically with the products of the equation
(\ref{2.14}) solutions. The probability density of the last
solutions is given by the integrand (\ref{2.404}).

The relations (\ref{11.5}), (\ref{2.6}), (\ref{2.10}), (\ref{2.13}),
(\ref{2.403}), (\ref{2.4031}) imply
\begin{eqnarray}
\label{2.4041} (\gamma_{l,\dot{l}}^{k_{1}} (\sigma^{0})
\gamma_{l,\dot{l}}^{k_{2}} (\sigma^{0}))_{m\dot{m}, n\dot{n}} = -
\delta_{k_{1}k_{2}} \delta_{mn} \delta_{\dot{m} \dot{n}} +
\sum_{k_{3}\, =\, 1}^{3}
(- 1)^{2\dot{l}}(i\epsilon^{k_{1}k_{2}k_{3}})^{2l + 2\dot{l}}\times \nonumber \\
\left( (\alpha_{l,\dot{l}} (0)\gamma_{l,\dot{l}}^{0} (\sigma^{0})
\gamma_{l,\dot{l}}^{k_{3}} (\sigma^{0}))_{m\dot{m}, n\dot{n}} -
(\beta_{l,\dot{l}} (0)\gamma_{l,\dot{l}}^{0} (\sigma^{0})
\gamma_{l,\dot{l}}^{k_{3}} (\sigma^{0}))_{m\dot{m}, n\dot{n}}
\right), \nonumber \\ m,n = 1,...,4l + 4,\, \, \dot{m}, \dot{n} =
1,...,2\dot{l} + 1,\, \, k_{1},k_{2} = 1,2,3.
\end{eqnarray}
For an odd integer $2l + 2\dot{l}$ the relations (\ref{2.403}),
(\ref{2.4031}), (\ref{2.4041}) imply
\begin{eqnarray}
\label{2.4042} (\gamma_{l,\dot{l}}^{\mu} (\sigma^{0})
\gamma_{l,\dot{l}}^{\nu} (\sigma^{0}))_{m\dot{m}, n\dot{n}} +
(\gamma_{l,\dot{l}}^{\nu} (\sigma^{0})
\gamma_{l,\dot{l}}^{\mu}(\sigma^{0}))_{m\dot{m}, n\dot{n}} =
2\eta^{\mu \nu} \delta_{m\dot{m}, n\dot{n}}, \nonumber \\ \mu, \nu =
0,...,3,\, \, m,n = 1,...,4l + 2,\, \,\dot{m}, \dot{n} =
1,...,2\dot{l} + 1.
\end{eqnarray}

Let us define the representation of the group $SL(2,{\bf C})$ in the
Lorentz group
\begin{equation}
\label{2.18} \sum_{\mu, \nu \, =\, 0}^{3} \Lambda_{\nu}^{\mu} (A)
x^{\nu}\sigma^{\mu} = A\tilde{x} A^{\ast}.
\end{equation}
We define also the $(4l + 2)(2\dot{l} + 1)$ - dimensional
representation of the group $SL(2,{\bf C})$
\begin{eqnarray}
\label{2.19} (S_{l,\dot{l}}(A))_{m\dot{m}, n\dot{n}} = t_{m - l -
1,n - l - 1}^{l}(A)t_{\dot{m} - \dot{l} - 1,\dot{n} - \dot{l} -
1}^{\dot{l}}(\bar{A}), \nonumber \\  m,n = 1,...,2l + 1,\, \,
\dot{m}, \dot{n} = 1,...,2\dot{l} + 1; \nonumber
\\ (S_{l,\dot{l}}(A))_{m\dot{m}, n\dot{n}} = t_{m - 3l - 2,n - 3l -
2}^{l}((A^{\ast})^{- 1})t_{\dot{m} - \dot{l} - 1,\dot{n} - \dot{l} -
1}^{\dot{l}}((A^{T})^{- 1}), \nonumber \\ m,n = 2l + 2,...,4l + 2,\,
\, \dot{m}, \dot{n} = 1,...,2\dot{l} + 1.
\end{eqnarray}
The other matrix elements are equal to zero. The definitions
(\ref{2.13}), (\ref{2.19}) imply
\begin{equation}
\label{2.20} S_{l,\dot{l}}(A)\alpha_{l,\dot{l}}(\mu^{2})
S_{l,\dot{l}}(A^{- 1}) = \alpha_{l,\dot{l}} (\mu^{2}), \, \,
S_{l,\dot{l}}(A)\beta_{l,\dot{l}}(\mu^{2}) S_{l,\dot{l}}(A^{- 1}) =
\beta_{l,\dot{l}} (\mu^{2}).
\end{equation}
Let the functions $\psi_{m\dot{m}} (x)$, $m = 1,...,4l + 2$,
$\dot{m} = 1,...,2\dot{l} + 1$ be the solutions of the equation
(\ref{2.14}). The relations (\ref{2.20}) imply that the functions
\begin{equation}
\label{2.21} \xi_{m\dot{m}} (\tilde{x}) = \sum_{n\, =\, 1}^{4l + 2}
\sum_{\dot{n} \, =\, 1}^{2\dot{l} + 1} (S_{l,\dot{l}}(A))_{m\dot{m},
n\dot{n}} \psi_{n\dot{n}} (A^{- 1}\tilde{x} (A^{\ast})^{- 1})
\end{equation}
are the solutions of the equation
\begin{equation}
\label{2.22} \sum_{n\, =\, 1}^{4l + 2} \sum_{\dot{n} \, =\,
1}^{2\dot{l} + 1} \left( \sum_{\nu \, = \, 0}^{3}
(\gamma_{l,\dot{l}}^{\nu} (A))_{m\dot{m}, n\dot{n}} \left( -
i\frac{\partial}{\partial x^{\nu}}\right) + (\beta_{l,\dot{l}}
(\mu^{2}))_{m\dot{m}, n\dot{n}} \right) \xi_{n\dot{n}} (x) = 0,
\end{equation}
\begin{equation}
\label{2.23} \gamma_{l,\dot{l}}^{\mu} (A) = \sum_{\nu \, =\, 0}^{3}
\Lambda_{\nu}^{\mu} (A)S_{l,\dot{l}} (A)\gamma_{l,\dot{l}}^{\nu}
(\sigma^{0}) S_{l,\dot{l}} (A^{- 1})
\end{equation}
for any matrix $A \in SL(2,{\bf C})$. The definition (\ref{2.23})
implies
\begin{equation}
\label{2.24} \gamma_{l,\dot{l}}^{\mu} (AB) = \sum_{\nu \, =\, 0}^{3}
\Lambda_{\nu}^{\mu} (A)S_{l,\dot{l}} (A)\gamma_{l,\dot{l}}^{\nu} (B)
S_{l,\dot{l}} (A^{- 1})
\end{equation}
for any matrices $A,B \in SL(2,{\bf C})$. By changing the coordinate
system we change the matrix $\gamma_{l,\dot{l}}^{\nu} (\sigma^{0})$
in the equation (\ref{2.14}) for the matrix (\ref{2.23}). The
solutions of the equation (\ref{2.14}) are transformed to the
solutions (\ref{2.21}) of the equation (\ref{2.22}). It is valid for
all non - negative half - integers $l,\dot{l}$. Due to the book
(\cite{6}, relation (1 - 43)) for any matrix $A \in SL(2,{\bf C})$
\begin{equation}
\label{2.25} \gamma_{\frac{1}{2}, 0}^{\mu} (A) =
\gamma_{\frac{1}{2}, 0}^{\mu} (\sigma^{0})
\end{equation}
Hence the equation (\ref{2.14}) for $l = 1/2$, $\dot{l} = 0$ is
covariant under the group $SL(2,{\bf C})$. The relation (\ref{2.25})
picks out Dirac equation.

For an odd integer $2l + 2\dot{l}$ the relations (\ref{2.4042}),
(\ref{2.23}) imply
\begin{eqnarray}
\label{2.4043} (\gamma_{l,\dot{l}}^{\mu} (A)
\gamma_{l,\dot{l}}^{\nu} (A))_{m\dot{m}, n\dot{n}} +
(\gamma_{l,\dot{l}}^{\nu} (A)
\gamma_{l,\dot{l}}^{\mu}(A))_{m\dot{m}, n\dot{n}} = 2\eta^{\mu \nu}
\delta_{m\dot{m}, n\dot{n}}, \nonumber \\ \mu, \nu = 0,...,3,\, \,
m,n = 1,...,4l + 2,\, \,\dot{m}, \dot{n} = 1,...,2\dot{l} + 1
\end{eqnarray}
for any matrix $A \in SL(2,{\bf C})$.

Let the functions $\xi_{m\dot{m}} (x)$ be the solutions of the
equation (\ref{2.22}). Let us introduce the distributions
\begin{equation}
\label{2.30} f_{m\dot{m}} (x) = \theta (x^{0})\xi_{m\dot{m}} (x),\,
\, \theta (x) = \left\{ {1, \hskip 0,5cm x \geq 0,} \atop {0, \hskip
0,5cm x < 0.} \right.
\end{equation}
The equation (\ref{2.22}) implies
\begin{eqnarray}
\label{2.32} \sum_{n\, =\, 1}^{4l + 2} \sum_{\dot{n} \, =\,
1}^{2\dot{l} + 1}  \left( \sum_{\nu \, = \, 0}^{3}
(\gamma_{l,\dot{l}}^{\nu} (A))_{m\dot{m}, n\dot{n}} \left( -
i\frac{\partial}{\partial x^{\nu}}\right) + (\beta_{l,\dot{l}}
(\mu^{2}))_{m\dot{m}, n\dot{n}} \right) f_{n\dot{n}} (x) = \nonumber
\\ - i\delta (x^{0})f^{0}_{m\dot{m}} (+ 0,{\bf x}), \\
f^{0}_{m\dot{m}} (+ 0,{\bf x}) = \sum_{n\, =\, 1}^{4l + 2}
\sum_{\dot{n} \, =\, 1}^{2\dot{l} + 1} (\gamma_{l,\dot{l}}^{0}
(A))_{m\dot{m}, n\dot{n}} f_{n\dot{n}} (+ 0,{\bf x}). \nonumber
\end{eqnarray}

Let a distribution $e_{\mu_{1}^{2},..., \mu_{n}^{2}} (x) \in
S^{\prime} ({\bf R}^{4})$ with support in the closed upper light
cone satisfy the equation
\begin{equation}
\label{2.33} \left( \prod_{i\, =\, 1}^{n} (- (\partial_{x},
\partial_{x}) - \mu_{i}^{2}) \right) e_{\mu_{1}^{2},..., \mu_{n}^{2}} (x) =
\delta (x).
\end{equation}
We prove the uniqueness of the equation (\ref{2.33}) solution in the
class of the distributions with supports in the closed upper light
cone. Let the equation (\ref{2.33}) have two solutions
$e_{\mu_{1}^{1},..., \mu_{n}^{2}}^{(1)} (x)$, $e_{\mu_{1}^{2},...,
\mu_{n}^{2}}^{(2)} (x)$. Since its supports lie in the closed upper
light cone the convolution is defined. Now the convolution
commutativity implies these distribution coincidence:
\begin{eqnarray}
\label{2.331} e_{\mu_{1}^{2},..., \mu_{n}^{2}}^{(2)} (x) = \left(
\prod_{i\, =\, 1}^{n} (- (\partial_{x},
\partial_{x}) - \mu_{i}^{2}) \right) \int d^{4}y e_{\mu_{1}^{2},...,
\mu_{n}^{2}}^{(1)} (x - y) e_{\mu_{1}^{2},..., \mu_{n}^{2}}^{(2)}
(y) = \nonumber \\ \left( \prod_{i\, =\, 1}^{n} (- (\partial_{x},
\partial_{x}) - \mu_{i}^{2}) \right) \int d^{4}y e_{\mu_{1}^{2},...,
\mu_{n}^{2}}^{(2)} (x - y) e_{\mu_{1}^{2},..., \mu_{n}^{2}}^{(1)}
(y) = e_{\mu_{1}^{2},..., \mu_{n}^{2}}^{(1)} (x).
\end{eqnarray}
Due to the book (\cite{3}, Section 30)
\begin{equation}
\label{2.34} e_{0}(x) = -\, (2\pi)^{- 1} \theta (x^{0})\delta
((x,x)),\, \, e_{0,0}(x) = (8\pi)^{- 1}\theta (x^{0})\theta ((x,x)).
\end{equation}
The uniqueness of the equation (\ref{2.33}) solution in the class of
the distributions with supports in the closed upper light cone
implies
\begin{equation}
\label{12.42} - (\partial_{x}, \partial_{x}) e_{0,0}(x) = e_{0}(x).
\end{equation}
In view of the second definition (\ref{2.30})
\begin{equation}
\label{2.35} (\partial_{x}, \partial_{x}) (\theta (x^{0})\theta
((x,x))(x,x)^{n}) = 4n(n + 1)\theta (x^{0})\theta ((x,x))(x,x)^{n -
1},\, \, n = 1,2,....
\end{equation}
Due to the relations (\ref{2.34}), (\ref{2.35}) the distribution
$e_{0,...,0}(x)$ with $n$ zeros has the form
\begin{equation}
\label{2.36} e_{0,...,0}(x) = (- 1)^{n}(2\pi 4^{n - 1}(n - 2)!(n -
1)!)^{- 1}\theta (x^{0})\theta ((x,x))(x,x)^{n - 2},\, \, n =
2,3,....
\end{equation}
Let us prove the following equality
\begin{eqnarray}
\label{2.361} e_{\mu_{1}^{2}}(x) =  -\, (2\pi)^{- 1} \theta
(x^{0})\delta ((x,x)) + \nonumber \\ \theta (x^{0})\theta ((x,x))
\sum_{n\, =\, 1}^{\infty} \mu_{1}^{2n} (- 1)^{n + 1}(2\pi 4^{n}(n -
1)!n!)^{- 1} (x,x)^{n - 1}.
\end{eqnarray}
By making use of the relations (\ref{2.34}), (\ref{12.42}),
(\ref{2.35}) it is possible to prove that the right - hand side of
the equality (\ref{2.361}) satisfies the equation (\ref{2.33}) for
$n = 1$. The support of the right - hand side of the equality
(\ref{2.361}) lies in the closed upper light cone. The solution of
the equation (\ref{2.33}) is unique in the class of the
distributions with supports in the closed upper light cone.

Let us prove
\begin{eqnarray}
\label{2.37} e_{\mu_{1}^{2},..., \mu_{n}^{2}} (x) = \lim_{\epsilon
\rightarrow + 0} (2\pi)^{- 4} \int d^{4}p\exp \{ - i(p,x)\}
\prod_{j\, =\, 1}^{n} ((p^{0} + i\epsilon )^{2} - |{\bf p}|^{2} -
\mu_{j}^{2} )^{- 1}.
\end{eqnarray}
The integral (\ref{2.37}) is the solution of the equation
(\ref{2.33}). By making the shift of the integration path in the
right - hand side of the equality (\ref{2.37}) we obtain that the
distribution (\ref{2.37}) is equal to zero for $x^{0} < 0$. The
distribution (\ref{2.37}) is Lorentz invariant. Hence its support
lies in the closed upper light cone. Now the uniqueness of the
distribution $e_{\mu_{1}^{2},..., \mu_{n}^{2}} (x)$ implies the
equality (\ref{2.37}). The equality (\ref{1.6}) is the particular
case of the equality (\ref{2.37}).

The relations (\ref{2.15}), (\ref{2.401}), (\ref{2.4043}) imply the
relation
\begin{eqnarray}
\label{2.38} \sum_{p\, =\, 1}^{4l + 2} \sum_{\dot{p} \, =\,
1}^{2\dot{l} + 1} \left( \sum_{\nu \, =\, 0}^{3}
(\gamma_{l,\dot{l}}^{\nu} (A))_{m\dot{m}, p\dot{p}} \left( -
i\frac{\partial}{\partial x^{\nu}} \right) + (\beta_{l,\dot{l}}
(\mu^{2}))_{m\dot{m}, p\dot{p}}\right) \times \nonumber
\\ \left( \sum_{\nu \, =\, 0}^{3} (\gamma_{l,\dot{l}}^{\nu}
(A))_{p\dot{p}, n\dot{n}} \left( - i\frac{\partial}{\partial
x^{\nu}} \right) - (\alpha_{l,\dot{l}} (\mu^{2}))_{p\dot{p},
n\dot{n}} \right) = ( - (\partial_{x},
\partial_{x}) - \mu^{2}) \delta_{mn} \delta_{\dot{m} \dot{n}},
\nonumber \\
m,n = 1,...,4l + 2,\, \, \dot{m}, \dot{n} = 1,...,2\dot{l} + 1,
\end{eqnarray}
for an odd integer $2l + 2\dot{l}$. The relations (\ref{2.33}),
(\ref{2.38}) imply that the solution of the equation (\ref{2.32})
has the form
\begin{eqnarray}
\label{2.39} f_{m\dot{m}} (x) = \sum_{n\, =\, 1}^{4l + 2}
\sum_{\dot{n} \, =\, 1}^{2\dot{l} + 1} \left( \sum_{\nu \, =\,
0}^{3} (\gamma_{l,\dot{l}}^{\nu} (A))_{m\dot{m}, n\dot{n}} \left( -
i\frac{\partial}{\partial x^{\nu}} \right) - (\alpha_{l,\dot{l}}
(\mu^{2}))_{m\dot{m}, n\dot{n}} \right) \times \nonumber
\\ \left( - i\int d^{4}ye_{\mu^{2}} (x - y)\delta (y^{0})
f^{0}_{n\dot{n}} (+ 0,{\bf y})\right),
\end{eqnarray}
$m = 1,...,4l + 2,\, \dot{m} = 1,...,2\dot{l} + 1$. For the solution
of the equation (\ref{2.22}) in the domain $x^{0} < 0$ it is
sufficient to use the distribution $- e_{\mu^{2}} (- x)$ in the
relation (\ref{2.39}).

We suppose that the smooth function $f_{n\dot{n}} (+ 0,{\bf x})$ is
rapidly decreasing at the infinity. By shifting the integration path
in the integral (\ref{2.37}) we have
\begin{eqnarray}
\label{2.40}  \int d^{4}ye_{\mu^{2}} (x - y)\delta (y^{0})
f_{n\dot{n}} (+ 0,{\bf y}) = \nonumber \\ (2\pi)^{- 4} \int
d^{4}p\exp \{ x^{0} - i(p,x)\} ((p^{0} + i)^{2} - |{\bf p}|^{2} -
\mu^{2})^{- 1}
\tilde{f}_{n\dot{n}} (+ 0,\cdot ) ({\bf p}), \nonumber \\
\tilde{f}_{m\dot{m}} (+ 0,\cdot ) ({\bf p}) = \int d^{3}{\bf x} \exp
\{ - i\sum_{k\, =\, 1}^{3} p^{k}x^{k}\} f_{m\dot{m}} (+ 0,{\bf x}).
\end{eqnarray}
The integral with respect to $p^{0}$ may be easily calculated. For
$x^{0} > 0$ and $ \tilde{f}_{m\dot{m}} (+ 0,\cdot )({\bf p}) =
f_{m\dot{m}} \delta ({\bf p} - {\bf q}) $ the functions
(\ref{2.39}), (\ref{2.40}) are not the eigenfunctions of the
differential operator $ - i\partial /\partial x^{0} $ and are the
eigenfunctions of the differential operator $\left( - i\partial
/\partial x^{0} \right)^{2}$.

\section{Relativistic quantum Coulomb law}
\setcounter{equation}{0}

We consider at first the relativistic Coulomb law in the classical
mechanics. The relativistic Coulomb law is the particular case of
the relativistic Newton second law
\begin{eqnarray}
\label{1.17} mc\frac{dt}{ds} \frac{d}{dt} \left( \frac{dt}{ds}
\frac{dx^{\mu}}{dt} \right) + \frac{q}{c} \sum_{k\, =\, 0}^{N}
\sum_{\alpha_{1}, ..., \alpha_{k} \, =\, 0}^{3} \eta^{\mu \mu}
F_{\mu \alpha_{1} \cdots \alpha_{k}}(x)\frac{dt}{ds}
\frac{dx^{\alpha_{1}}}{dt} \cdots \frac{dt}{ds}
\frac{dx^{\alpha_{k}}}{dt} = 0, \nonumber \\ \frac{dt}{ds} = c^{-
1}\left( 1 - c^{- 2}\Bigl| \frac{d{\bf x}}{dt}\Bigr|^{2}\right)^{-
1/2}
\end{eqnarray}
where $x^{0} = ct$, $\mu = 0,...,3$. In the equation (\ref{1.17})
the force is the polynomial of the speed. For an infinite series of
the speed it is necessary to define the series convergence. The
second relation (\ref{1.17}) implies the identities
\begin{equation}
\label{1.19} \sum_{\alpha \, =\, 0}^{3} \eta_{\alpha \alpha} \left(
\frac{dt}{ds} \frac{dx^{\alpha}}{dt} \right)^{2} = 1, \, \,
\sum_{\alpha \, =\, 0}^{3} \eta_{\alpha \alpha} \frac{dt}{ds}
\frac{dx^{\alpha}}{dt} \frac{dt}{ds} \frac{d}{dt} \left(
\frac{dt}{ds} \frac{dx^{\alpha}}{dt} \right) = 0.
\end{equation}
The equation (\ref{1.17}) and the second identity (\ref{1.19}) imply
\begin{equation}
\label{1.20} \sum_{k\, =\, 0}^{N} \sum_{\alpha_{1}, ..., \alpha_{k +
1} \, =\, 0}^{3} F_{\alpha_{1} \cdots \alpha_{k + 1}}(x)
\frac{dt}{ds} \frac{dx^{\alpha_{1}}}{dt} \cdots \frac{dt}{ds}
\frac{dx^{\alpha_{k + 1}}}{dt} = 0.
\end{equation}
Let the functions $F_{\alpha_{1} \cdots \alpha_{k + 1}}(x)$ satisfy
the equation (\ref{1.20}). Then three equations (\ref{1.17}) for
$\mu = 1,2,3$ are independent
\begin{eqnarray}
\label{1.21} m\frac{d}{dt} \left( (1 - c^{- 2}|{\bf v}|^{2})^{-
1/2}v^{i}\right) - q\sum_{k\, =\, 0}^{N} c^{- k}(1 - c^{- 2}|{\bf
v}|^{2})^{- \frac{k - 1}{2}} \times \nonumber \\ \sum_{\alpha_{1},
..., \alpha_{k} \, =\, 0}^{3} F_{i\alpha_{1} \cdots
\alpha_{k}}(x)\frac{dx^{\alpha_{1}}}{dt} \cdots
\frac{dx^{\alpha_{k}}}{dt} = 0, \, \, v^{i} = \frac{dx^{i}}{dt},\,
\, i = 1,2,3.
\end{eqnarray}
The following lemma is proved in the paper \cite{12}.

\noindent {\bf Lemma}. {\it Let there exist a Lagrange function}
$L({\bf x},{\bf v},t)$ {\it such that for any world line},
$x^{\mu}(t)$, $x^{0}(t) = ct$, {\it and for any} $i = 1,2,3$ {\it
the relation}
\begin{eqnarray}
\label{1.22} \frac{d}{dt} \frac{\partial L}{\partial v^{i}} -
\frac{\partial L}{\partial x^{i}} = m\frac{d}{dt} \left( (1 - c^{-
2}|{\bf v}|^{2})^{- 1/2}v^{i}\right) - \nonumber \\ q\sum_{k\, =\,
0}^{N} c^{- k}(1 - c^{- 2}|{\bf v}|^{2})^{- \frac{k - 1}{2}}
\sum_{\alpha_{1}, ..., \alpha_{k} \, =\, 0}^{3}F_{i\alpha_{1} \cdots
\alpha_{k}}(x)\frac{dx^{\alpha_{1}}}{dt} \cdots
\frac{dx^{\alpha_{k}}}{dt}
\end{eqnarray}
{\it holds. Then the Lagrange function has the form}
\begin{equation}
\label{1.23} L({\bf x},{\bf v},t) = - mc^{2}(1 - c^{- 2}|{\bf
v}|^{2})^{1/2} + \frac{q}{c} \sum_{i\, =\, 1}^{3} A_{i}({\bf
x},t)v^{i} + qA_{0}({\bf x},t)
\end{equation}
{\it and the coefficients in the equations} (\ref{1.21}) {\it are}
\begin{equation}
\label{1.24} F_{i\alpha_{1} \cdots \alpha_{k}}(x) = 0,\, \, k \neq
1,\, i = 1,2,3,\, \alpha_{1}, ...,\alpha_{k} = 0,...,3,
\end{equation}
\begin{eqnarray}
\label{1.25} F_{ij}(x) = \frac{\partial A_{j}({\bf x},t)}{\partial
x^{i}} - \frac{\partial A_{i}({\bf x},t)}{\partial x^{j}},\, \, i,j
= 1,2,3, \nonumber
\\ F_{i0}(x) = \frac{\partial A_{0}({\bf x},t)}{\partial
x^{i}} - \frac{1}{c} \frac{\partial A_{i}({\bf x},t)}{\partial t},\,
\, i = 1,2,3.
\end{eqnarray}
We define
\begin{equation}
\label{1.26} F_{00} = 0,\, \, F_{0i} = - F_{i0},\, \, i = 1,2,3.
\end{equation}
Then the identity
\begin{equation}
\label{1.27} \sum_{\alpha, \beta \, =\, 0}^{3} F_{\alpha
\beta}(x)\frac{dt}{ds} \frac{dx^{\alpha}}{dt} \frac{dt}{ds}
\frac{dx^{\beta}}{dt} = 0
\end{equation}
similar to the identity (\ref{1.20}) holds. By making use of the
second identity (\ref{1.19}) and the relations (\ref{1.25}) -
(\ref{1.27}) we can rewrite the equation (\ref{1.21}) with the
coefficients (\ref{1.24}), (\ref{1.25}) as the equation with Lorentz
force
\begin{eqnarray}
\label{1.28} mc\frac{dt}{ds} \frac{d}{dt} \left( \frac{dt}{ds}
\frac{dx^{\mu}}{dt} \right) = - \frac{q}{c} \eta^{\mu \mu} \sum_{\nu
\, =\, 0}^{3} F_{\mu \nu}(x)\frac{dt}{ds} \frac{dx^{\nu}}{dt},
\nonumber \\ F_{\mu \nu}(x) = \frac{\partial A_{\nu}({\bf
x},t)}{\partial x^{\mu}} - \frac{\partial A_{\mu}({\bf
x},t)}{\partial x^{\nu}}, \, \, \mu, \nu = 0,...,3.
\end{eqnarray}
The interaction is defined by the product of the charge $q$ and the
vector potential $A_{\mu}({\bf x},t)$ of the external field.

The relativistic Coulomb law is given by Lorentz covariant equations
describing the electromagnetic interaction of two particles with the
charges $q_{k}$, $k = 1,2$,
\begin{eqnarray}
\label{12.1} m_{k}c\frac{dt}{ds_{k}} \frac{d}{dt} \left(
\frac{dt}{ds_{k}} \frac{dx_{k}^{\mu}}{dt} \right) = -
\frac{q_{k}}{c} \eta^{\mu \mu} \sum_{\nu \, =\, 0}^{3} F_{j;\mu
\nu}(x_{k},x_{j})\frac{dt}{ds_{k}} \frac{dx_{k}^{\nu}}{dt},
\nonumber \\ \frac{dt}{ds_{k}} = c^{- 1}\left( 1 - c^{- 2}\Bigl|
\frac{d{\bf x}_{k}}{dt}\Bigr|^{2}\right)^{- 1/2},
\end{eqnarray}
\begin{equation}
\label{12.2} F_{j;\mu \nu}(x_{k},x_{j}) = \frac{\partial
A_{j;\nu}(x_{k},x_{j})}{\partial x_{k}^{\mu}} - \frac{\partial
A_{j;\mu}(x_{k},x_{j})}{\partial x_{k}^{\nu}},
\end{equation}
for any permutation $k,j$ of the integers $1,2$. The world lines
$x_{k}(t)$, $k = 1,2$, satisfy the condition  $x_{k}^{0}(t) = ct$
where $c$ is the speed of light. Li\'enard - Wiechert vector
potentials are given by the following relations
\begin{eqnarray}
\label{12.4}  A_{j; \mu}(x_{k},x_{j}) = -\, 4\pi q_{j}K \sum_{\nu \,
= \, 0}^{3} \eta_{\mu \nu} \int dt e_{0}(x_{k} - x_{j}(t))
\frac{dx_{j}^{\nu }(t)}{dt} = \nonumber \\
q_{j}K \eta_{\mu \mu} \left( \frac{d}{dt(0)}
x_{j}^{\mu}(t(0))\right) \left( c|{\bf x}_{k} - {\bf x}_{j}(t(0))| -
\sum_{i\, =\, 1}^{3} (x_{k}^{i} - x_{j}^{i}(t(0))) \frac{d}{dt(0)}
x_{j}^{i}(t(0))\right)^{- 1}, \nonumber \\ x_{k}^{0} - ct(0) = |{\bf
x}_{k} - {\bf x}_{j}(t(0))|
\end{eqnarray}
where $e_{0}(x)$ is the first distribution (\ref{2.34}). By making
change of the integration variable
\begin{equation}
\label{12.17} x_{k}^{0} - ct(r) = (|{\bf x}_{k} - {\bf
x}_{j}(t(r))|^{2} + r)^{1/2}
\end{equation}
it is easy to prove the relation (\ref{12.4}). For $r = 0$ the
relation (\ref{12.17}) coincides with the last relation
(\ref{12.4}).

For a world line $x_{j}^{\mu}(t)$ we define the vector
\begin{eqnarray}
\label{12.10} J^{\mu}(x,x_{j}) = \int dt \delta (x - x_{j}(t))
\frac{dx_{j}^{\mu }(t)}{dt} = \nonumber \\ \left( \frac{d}{dx^{0}}
x_{j}^{\mu} \left( c^{- 1}x^{0} \right) \right) \delta \left( {\bf
x} - {\bf x}_{j} \left( c^{- 1} x^{0} \right) \right),\, \mu =
0,...,3.
\end{eqnarray}
The world line $x_{j}^{\mu}(t)$ satisfies the condition
$x_{j}^{0}(t) = ct$. Hence the definition (\ref{12.10}) implies the
continuity equation
\begin{eqnarray}
\label{12.12} \frac{\partial}{\partial x^{0}} J^{0}(x,x_{j}) = -
\sum_{i\, =\, 1}^{3} \left( \frac{d}{dx^{0}} x_{j}^{i} \left( c^{-
1}x^{0} \right) \right) \frac{\partial}{\partial x^{i}} \delta
\left( {\bf x} - {\bf
x}_{j} \left( c^{- 1} x_{k}^{0} \right) \right), \nonumber \\
\frac{\partial}{\partial x^{i}} J^{i}(x,x_{j}) = \left(
\frac{d}{dx^{0}} x_{j}^{i} \left( c^{- 1}x^{0} \right) \right)
\frac{\partial}{\partial x^{i}} \delta \left( {\bf x} - {\bf x}_{j}
\left( c^{- 1} x^{0} \right) \right),\, i = 1,2,3, \nonumber \\
\sum_{\mu \, =\, 0}^{3} \frac{\partial}{\partial x^{\mu}}
J^{\mu}(x,x_{j}) = 0.
\end{eqnarray}
The world line $x_{j}^{\mu}(t)$, $x_{j}^{0}= ct$, is called time
like, if it satisfies the condition
$$
\Bigl| \frac{d{\bf x}_{j}(t)}{dt}\Bigr| < c,
$$
For the time like world line $x_{j}^{\mu}(t)$ the supports of the
distributions (\ref{12.10}) lie in the closed upper light cone and
\begin{equation}
\label{12.14} \int dt e_{0}(x - x_{j}(t)) \frac{dx_{j}^{\mu
}(t)}{dt} = \int d^{4}y e_{0}(x - y) \int dt \delta (y - x_{j}(t))
\frac{dx_{j}^{\mu}(t)}{dt}.
\end{equation}
The supports of both sides of the equality (\ref{12.14}) lie in the
closed upper light cone. The distribution $ e_{0}(x)$ satisfies the
equation (\ref{2.33}). Hence both sides of the equality
(\ref{12.14}) satisfy the equation
$$
(- (\partial_{x},\partial_{x})) f^{\mu}(x) = J^{\mu}(x,x_{j}).
$$
The difference of two solutions of this equation is a solution of
the equation (\ref{2.33}). Since the solution of this equation
similar to the solution of the equation (\ref{2.33}) is unique in
the class of the distributions with supports in the closed upper
light cone both sides of the equality (\ref{12.14}) coincide.

The relations (\ref{12.12}), (\ref{12.14}) imply the gauge condition
for the vector potential (\ref{12.4})
\begin{equation}
\label{12.16} \sum_{\mu \, =\, 0}^{3} \eta^{\mu \mu}
\frac{\partial}{\partial x^{\mu}} A_{j;\mu}(x,x_{j}) = 0.
\end{equation}
Due to relation (\ref{12.16}) the tensor of the electromagnetic
field (\ref{12.2}), (\ref{12.4}) satisfies Maxwell equations with
the current $4\pi q_{j}K \eta_{\mu \mu} J^{\mu}(x_{k},x_{j})$.

In contrast to the equation (\ref{1.3}) the equations (\ref{12.1}) -
(\ref{12.4}) satisfy the causality condition. The vector potential
(\ref{12.4}) depends only on the world line points $x_{j}^{\mu}(t)$
lying in the closed lower light cone with the vertex at the point
$x_{k}$.

In the equations (\ref{12.1}) - (\ref{12.4}) the electromagnetic
interaction is defined by Li\'enard - Wiechert vector potentials.
Let us consider the interaction coefficients
\begin{eqnarray}
\label{12.6} A_{\mu \nu}^{(jk)} (x) = K_{jk;0}Q_{\mu \nu}^{0}
(\partial_{x}) e_{0,0}(x) + K_{jk;1}Q_{\mu
\nu}^{1} (\partial_{x}) e_{0,0}(x), \nonumber \\
Q_{\mu \nu}^{0} (\partial_{x}) = - \eta_{\mu \nu} (\partial_{x},
\partial_{x}), \, \, Q_{\mu \nu}^{1} (\partial_{x}) =
4\frac{\partial}{\partial x^{\mu}} \frac{\partial}{\partial x^{\nu}}
- \eta_{\mu \nu} (\partial_{x}, \partial_{x}).
\end{eqnarray}
The interaction coefficient (\ref{12.6}) is defined by two constants
$K_{jk;0}$, $K_{jk;1}$. Due to the relation (\ref{12.42}) for
$K_{jk;1} = 0$ the interaction coefficient (\ref{12.6}) is equal to
$\eta_{\mu \nu} K_{jk;0}e_{0}(x)$. The polynomials  $Q_{\mu \nu}^{l}
(\partial_{x})$, $l = 0,1$, are connected to Minkowski metric
\begin{equation}
\label{12.7}  \eta_{\mu \nu} \sum_{\alpha, \beta \, = \, 0,...,3}
\eta^{\alpha \beta} Q_{\alpha \beta}^{0} (\partial_{x}) = 4Q_{\mu
\nu}^{0} (\partial_{x}), \, \, \eta_{\mu \nu} \sum_{\alpha, \beta \,
= \, 0,...,3} \eta^{\alpha \beta} Q_{\alpha \beta}^{1}
(\partial_{x}) = 0.
\end{equation}
We change in the equations (\ref{12.1}) - (\ref{12.4}) the
interaction coefficients $4\pi q_{k}q_{j}c^{- 1} K\eta_{\mu \nu}
e_{0}(x)$ for the interaction coefficients (\ref{12.6}). The
relation (\ref{12.14}) is still valid when the distribution
$e_{0,0}(x)$ substitutes for the distribution $e_{0}(x)$. By making
use of the obtained relation and the relations (\ref{12.42}),
(\ref{12.12}), (\ref{12.6}) it is easy to prove the relation
\begin{equation}
\label{12.5} \sum_{\nu \, = \, 0}^{3}
\int dt A_{\mu \nu}^{(jk)}(x_{k} - x_{j}(t)) \frac{dx_{j}^{\nu }(t)}{dt}
= (K_{jk;0} + K_{jk;1})\sum_{\nu \, = \, 0}^{3} \eta_{\mu \nu}
\int dt e_{0}(x_{k} - x_{j}(t))\frac{dx_{j}^{\nu }(t)}{dt}.
\end{equation}
The constants $K_{jk;l}$, $l = 0,1$, of the interaction coefficient
(\ref{12.6}) are included in the vector potential (\ref{12.5}) as
the sum $K_{jk;0} + K_{jk;1}$.

Let us introduce the interaction coefficients into the equation
(\ref{2.32}). Let $j,k$ be the permutation of the numbers $1,2$. We
construct the equation for $j$ particle. Let us multiply the
equations (\ref{2.32}) for the particles $1$ and $2$. We change the
differential operator
\begin{equation}
\label{2.41} \left( - i\frac{\partial}{\partial
x_{1}^{\nu_{1}}}\right) \left( - i\frac{\partial}{\partial
x_{2}^{\nu_{2}}}\right)
\end{equation}
for the differential operator
\begin{equation}
\label{2.42} \left( - i\frac{\partial}{\partial
x_{1}^{\nu_{1}}}\right) \left( - i\frac{\partial}{\partial
x_{2}^{\nu_{2}}}\right) + A_{\nu_{1} \nu_{2}}^{(kj)} (x_{j} -
x_{k}).
\end{equation}
(We do not change in the equation (\ref{2.32}) the differential
operator $- i\partial_{\nu}$ for the differential operator $-
i\partial_{\nu} + qA_{\nu}(x)$). In the differential operators
(\ref{2.42}) the interaction coefficients $A_{\nu_{1}
\nu_{2}}^{(jk)} (x)$ are given by the relations (\ref{12.6}). For
any matrix $A \in SL(2,{\bf C})$ the equality
\begin{equation}
\label{2.43} A_{\nu_{1} \nu_{2}}^{(kj)} \left( \sum_{\mu \, =\,
0}^{3} \Lambda_{\mu}^{\lambda} (A^{- 1})x^{\mu} \right) =
\sum_{\mu_{1}, \mu_{2} \, =\, 0}^{3} \Lambda_{\nu_{1}}^{\mu_{1}}
(A)\Lambda_{\nu_{2}}^{\mu_{2}} (A)A_{\mu_{1} \mu_{2}}^{(kj)} (x)
\end{equation}
holds. The differential operator (\ref{2.42}) transformation is
similar to the differential operator (\ref{2.41}) transformation.
The support of the distribution (\ref{12.6}) lies in the closed
upper light cone. Hence the differential operator (\ref{2.42})
differs from the differential operator (\ref{2.41}) only for the
vector $x_{j} - x_{k}$ lying in the closed upper light cone. In
order to obtain an equation of the type (\ref{1.2}) we integrate our
equation with respect to the variable $x_{k}$
\begin{eqnarray}
\label{2.44} \sum_{n_{s} \, =\, 1,...,4l_{s} + 2,\, \, s\, =\, 1,2}
\sum_{\dot{n}_{s} \, =\, 1,...,2\dot{l}_{s} + 1,\, \, s\, =\, 1,2}
\int d^{4}x_{k} \nonumber \\ \Biggl( \prod_{s\, =\, 1}^{2} \left(
\sum_{\nu \, = \, 0}^{3} (\gamma_{l_{s},\dot{l}_{s}}^{\nu}
(A))_{m_{s}\dot{m}_{s}, n_{s}\dot{n}_{s}} \left( -
i\frac{\partial}{\partial x_{s}^{\nu}}\right) +
(\beta_{l_{s},\dot{l}_{s}} (\mu_{s}^{2}))_{m_{s}\dot{m}_{s},
n_{s}\dot{n}_{s}} \right) (f_{s})_{n_{s}\dot{n}_{s}} (x_{s}) +
\nonumber \\ \prod_{s\, =\, 1}^{2} (\gamma_{l_{s},\dot{l}_{s}}^{0}
(A))_{m_{s}\dot{m}_{s}, n_{s}\dot{n}_{s}} \delta
(x_{s}^{0})(f_{s})_{n_{s}\dot{n}_{s}} (+ 0,{\bf x}_{s}) + \nonumber
\\ \sum_{\nu_{1}, \nu_{2} \, =\, 0}^{3} A_{\nu_{1} \nu_{2}}^{(kj)}
(x_{j} - x_{k}) \prod_{s\, =\, 1}^{2}
(\gamma_{l_{s},\dot{l}_{s}}^{\nu_{s}} (A))_{m_{s}\dot{m}_{s},
n_{s}\dot{n}_{s}} (f_{s})_{n_{s}\dot{n}_{s}} (x_{s})\Biggr) = 0.
\end{eqnarray}
In the equations (\ref{12.4}) we integrate along the particle world
line the product of the interaction coefficient and the velocity
vector
$$
-\, 4\pi q_{j}K \sum_{\nu \, = \, 0}^{3} \eta_{\mu \nu} \int dt
e_{0}(x_{k} - x_{j}(t)) \frac{dx_{j}^{\nu }(t)}{dt}.
$$
(Due to the relation (\ref{12.5}) the interaction coefficient
(\ref{12.6}) gives a similar interaction term.) A particle has no a
world line in the quantum mechanics. The particle probability
density is given by the wave function. In the equations (\ref{2.44})
we integrate over the space the product of the same interaction
coefficient and the vector constructed by means of the wave function
$$
\int d^{4}x_{k} \sum_{\nu \, =\, 0}^{3}  A_{\mu \nu}^{(kj)} (x_{j} -
x_{k}) \sum_{n_{k} \, =\, 1}^{4l_{k} + 2} \sum_{\dot{n}_{k} \, =\,
1}^{2\dot{l}_{k} + 1} (\gamma_{l_{k},\dot{l}_{k}}^{\nu}
(A))_{m_{k}\dot{m}_{k}, n_{k}\dot{n}_{k}} (f_{k})_{n_{k}\dot{n}_{k}}
(x_{k}).
$$
The first and the second terms in the left - hand side of the
equation (\ref{2.44}) correspond to the multiplied equations
(\ref{2.32}) for the particles $1$ and $2$. The equation
(\ref{2.44}) transforms like two equations (\ref{2.32}).

The relations (\ref{12.6}) define the interaction coefficients
$A_{\nu_{1} \nu_{2}}^{(kj)} (x)$ in the equations (\ref{2.44}). The
degree of the homogeneous polynomials $Q_{\nu_{1} \nu_{2}}^{l}
(\partial_{x})$, $l = 0,1$, is $2l$. Hence the second relation
(\ref{2.34}) implies that the supports of the distributions
$Q_{\nu_{1} \nu_{2}}^{l} (\partial_{x}) e_{0,0}(x)$, $l = 0,1$, lie
on upper light cone boundary. Thus the equations (\ref{12.6}),
(\ref{2.44}) satisfy the causality condition. The interaction
propagates at the speed of light.

The definitions (\ref{2.37}), (\ref{12.6}) imply
\begin{eqnarray}
\label{2.62} A_{\mu \nu}^{(kj)} (x) = \lim_{\epsilon \rightarrow +
0} (K_{kj;0} + K_{kj;1})(2\pi)^{- 4} \int d^{4}p\exp \{ - i(p,x)\}
\times \nonumber \\ ((p^{0} + i\epsilon)^{2} - |{\bf p}|^{2})^{- 1}
\left( \eta_{\mu \nu} - \frac{4K_{kj;1}}{K_{kj;0} + K_{kj;1}}
\frac{p_{\mu} p_{\nu} }{(p^{0} + i\epsilon)^{2} - |{\bf p}|^{2}}
\right).
\end{eqnarray}
Due to (\cite{1}, Chapter 12, relation (12.91); \cite{7}, Chapter
III, relation (3.58)) the propagation function of the vector
particles in the Yang - Mills theory is
\begin{eqnarray}
\label{2.63} D_{\mu \nu}^{ab} (x) = \lim_{\epsilon \rightarrow + 0}
- \delta^{ab} (2\pi)^{- 4} \int d^{4}p\exp \{ - i(p,x)\} \times
\nonumber \\ ((p,p) + i\epsilon)^{- 1} \left( \eta_{\mu \nu} - (1 -
\alpha )\frac{p_{\mu} p_{\nu}}{(p,p) + i\epsilon }\right).
\end{eqnarray}
The number $\alpha$ is the consequence of the gauge condition for
the value $\partial_{\mu} A^{\mu}$. The choices $\alpha = 1$ and
$\alpha = 0$ are called the gauge conditions of Feynman and Landau
(\cite{1}, Section 12.2.2). The distribution (\ref{2.63}) differs
from the distribution (\ref{2.62}) in the rule of going around the
poles in the integral.

We have considered up to now that the interaction propagates at the
speed of light. Let the interaction propagate at the speed less or
equal to the speed of light. Hence the interaction coefficients
(\ref{12.6}) may be changed in the following way
\begin{equation}
\label{2.65} A_{\mu \nu}^{(kj)} (x) = \sum_{l\, = \, 0}^{1} \int
d\lambda_{1} d\lambda_{2} K_{kj;l}(\lambda_{1}, \lambda_{2}) Q_{\mu
\nu}^{l} (\partial_{x}) e_{\lambda_{1}^{2}, \lambda_{2}^{2}} (x).
\end{equation}
If the distributions $K_{kj;l}(\lambda_{1}, \lambda_{2}) =
K_{kj;l}\delta (\lambda_{1})\delta (\lambda_{2})$, $l = 1,2$, then
the interaction coefficient (\ref{2.65}) coincides with the
interaction coefficient (\ref{12.6}). The interaction coefficients
(\ref{2.65}) satisfy the covariance relation (\ref{2.43}). Changing
in the equations (\ref{2.44}) the interaction coefficients
(\ref{12.6}) for the interaction coefficients (\ref{2.65}) we obtain
in the general case the theory where the particles interact between
each other by means of the massive particles.

We consider the equations (\ref{2.44}), (\ref{12.6}) for the case
$l_{1} = l_{2} = 1/2$, $\dot{l}_{1} = \dot{l}_{2} = 0$. We suppose
that the constants $K_{12;l}$, $l = 0,1$, are small. If we neglect
the term containing the interaction coefficient $A_{\nu_{1}
\nu_{2}}^{(12)} (x)$, then the functions $(f_{2})_{m_{2}} (x_{2})$
are the solutions of the equation (\ref{2.32}). It is possible
simply to assume $K_{12;l} = 0$, $l = 0,1$. We do not know the
physical reason of the assumption $K_{12;l} = 0$, $K_{21;l} \neq 0$,
$l = 0,1$. However the differential equations (\ref{12.6}),
(\ref{2.44}) with these constants are mathematically self -
consistent.

Let the functions $(f_{2})_{m_{2}} (x_{2})$ be the solutions of the
equation (\ref{2.32}) and be given by the relation (\ref{2.39}). Let
the integral
\begin{eqnarray}
\label{3.18} \int d^{3}{\bf y}_{2} (f_{2}^{0})_{m_{2}} (+ 0,{\bf
y}_{2}) = \left\{ {\int d^{3}{\bf y}_{2} (f_{2})_{m_{2} + 2} (+
0,{\bf y}_{2}), \hskip 0,3cm m_{2} = 1,2,} \atop {\int d^{3}{\bf
y}_{2} (f_{2})_{m_{2} - 2} (+ 0,{\bf y}_{2}), \hskip 0,3cm m_{2} =
3,4} \right.
\end{eqnarray}
be not equal to zero for some index $m_{2}$. The equations
(\ref{2.32}), (\ref{2.44}) imply
\begin{eqnarray}
\label{3.19} \sum_{n_{1}\, =\, 1}^{4} \sum_{\nu \, =\, 0}^{3}
(\gamma^{\nu})_{m_{1} n_{1}} \left( - i\frac{\partial}{\partial
x_{1}^{\nu}}
+ B_{\nu}^{(21)}(x_{1}) \right) (f_{1})_{n_{1}} (x_{1}) + \nonumber \\
\sum_{n_{1}\, =\, 1}^{4} (\beta (\mu_{1}^{2}))_{m_{1} n_{1}}
(f_{1})_{n_{1}} (x_{1}) = - i \delta (x_{1}^{0}) (f_{1}^{0})_{m_{1}}
(+ 0,{\bf x}_{1}),
\end{eqnarray}
\begin{eqnarray}
\label{3.20} B_{\nu}^{(21)}(x_{1}) = i\left( \int d^{3}{\bf y}_{2}
(f_{2}^{0})_{m_{2}} (+ 0,{\bf y}_{2}) \right)^{- 1} \times \nonumber
\\ \sum_{\nu_{1} \, =\, 0}^{3} \int d^{4}x_{2} A_{\nu \nu_{1}}^{(21)}
(x_{1} - x_{2}) \sum_{{n_{2} \, =\, 1}}^{4} (\gamma^{\nu_{1}}
)_{m_{2} n_{2}} (f_{2})_{n_{2}} (x_{2}).
\end{eqnarray}
The equation (\ref{3.19}) is similar to the equation (\ref{1.2}).
The right - hand side of the equation (\ref{3.19}) defines Cauchy
problem. The first particle interacts with the field of second
particle. The vector potential of this field is given by the
equality (\ref{3.20}). The substitution of the first relation
(\ref{2.34}) and the relation (\ref{12.6}), $K_{21;1} = 0$, into the
vector potential (\ref{3.20}) yields the delayed vector potential of
second particle electromagnetic field
\begin{eqnarray}
\label{3.21} B_{\nu}^{(21)}(x_{1}) = \left( \int d^{3}{\bf y}_{2}
(f_{2}^{0})_{m_{2}} (+ 0,{\bf y}_{2}) \right)^{- 1} \sum_{\nu_{1} \,
=\, 0}^{3} \sum_{{n_{2} \, =\, 1}}^{4} \eta_{\nu \nu_{1}}
(\gamma^{\nu_{1}})_{m_{2} n_{2}} \times \nonumber
\\ (4\pi i)^{- 1}K_{21;0} \int d^{3}{\bf x}_{2} |{\bf x}_{1} - {\bf
x}_{2}|^{- 1} (f_{2})_{n_{2}} (x_{1}^{0} - |{\bf x}_{1} - {\bf
x}_{2}|,{\bf x}_{2}).
\end{eqnarray}
If we choose in the vector potential (\ref{3.21}) the functions
$(f_{2})_{m_{2}} (x_{2}) = (f_{2})_{m_{2}} \delta ({\bf x}_{2})$, we
get the vector potential of Coulomb type. The second particle moves
freely and the functions $(f_{2})_{m_{2}} (x_{2})$ are the solutions
of the equation (\ref{2.32}) given by the relation (\ref{2.39}). It
is possible to choose the initial functions $(f_{2})_{m_{2}} (+
0,{\bf x}_{2})$ only.

The following equality
\begin{equation}
\label{3.24} \int d^{4}y e_{\mu_{1}^{2}, ...,\mu_{n - 1}^{2}}(x -
y)e_{\mu_{n}^{2}}(y) = e_{\mu_{1}^{2}, ...,\mu_{n}^{2}}(x).
\end{equation}
holds. Both sides of the equality (\ref{3.24}) have supports in the
closed upper light cone and satisfy the equation (\ref{2.33}). The
solution of the equation (\ref{2.33}) is unique in the class of the
distributions with supports in the closed upper light cone. By
making use of the relations (\ref{2.39}), (\ref{12.6}), (\ref{3.24})
it is possible to rewrite the equality (\ref{3.20}) as
\begin{eqnarray}
\label{3.25} B_{\nu}^{(21)}(x) = \left( \int d^{3}{\bf y}_{2}
(f_{2}^{0})_{m_{2}} (+ 0,{\bf y}_{2}) \right)^{- 1} \times \nonumber
\\ \sum_{\nu_{1} \,
=\, 0}^{3} \sum_{{n_{2} \, =\, 1}}^{4} \left( \sum_{\nu_{2} \, =\,
0}^{3} (\gamma^{\nu_{1}} \gamma^{\nu_{2}})_{m_{2} n_{2}} \left( -
i\frac{\partial}{\partial x^{\nu_{2}}} \right) - (\gamma^{\nu_{1}}
\alpha (\mu_{2}^{2}))_{m_{2} n_{2}} \right) \times \nonumber
\\ \int d^{4}y(K_{21;0}Q_{\nu \nu_{1}}^{0}
(\partial_{x}) + K_{21;1}Q_{\nu \nu_{1}}^{1} (\partial_{x}))
e_{0,0,\mu_{2}^{2}}(x - y) \delta (y^{0}) (f_{2}^{0})_{n_{2}} (+
0,{\bf y}).
\end{eqnarray}
The relations (\ref{2.36}), (\ref{2.361}), (\ref{3.24}) imply the
equality
\begin{eqnarray}
\label{3.26} e_{0,0, \mu_{2}^{2}} (x) = \theta (x^{0})\theta ((x,x))
\sum_{n\, =\, 0}^{\infty} \mu_{2}^{2n} (- 1)^{n + 1}(2\pi 4^{n +
2}(n + 1)!(n + 2)!)^{- 1}(x,x)^{n + 1}.
\end{eqnarray}
By making use of the relations (\ref{2.15}), (\ref{12.6}),
(\ref{3.26}) we can rewrite in the open upper light cone the vector
potential (\ref{3.25}) with the functions $(f_{2})_{n_{2}} (+ 0,{\bf
y}) = (f_{2})_{n_{2}} \delta ({\bf y})$ as
\begin{equation}
\label{3.27} B_{\nu}^{(21)}(x) = B_{\nu}^{(21;0)} + B_{\nu}^{(21;1)}
(x)
\end{equation}
where the constant vector potential
\begin{equation}
\label{3.28} B_{\nu}^{(21;0)} = - (8\pi)^{- 1} K_{21;0} \left(
\sum_{{n_{2} \, =\, 1}}^{4} (\gamma^{0})_{m_{2} n_{2}}
(f_{2})_{n_{2}} \right)^{- 1} \sum_{{n_{2} \, =\, 1}}^{4} \eta_{\nu
\nu} (\gamma^{\nu} \gamma^{0} \beta (\mu_{2}^{2}))_{m_{2} n_{2}}
(f_{2})_{n_{2}}
\end{equation}
and the remainder vector potential
\begin{eqnarray}
\label{3.281} B_{\nu}^{(21;1)} (x) = \nonumber \\ \left( \int
d^{3}{\bf y}_{2} (f_{2}^{0})_{m_{2}} (+ 0,{\bf y}_{2}) \right)^{- 1}
\sum_{n\, =\, 1}^{\infty} \mu_{2}^{2n} (- 1)^{n +
1}(2\pi 4^{n + 2}(n + 1)!(n + 2)!)^{- 1} \times \nonumber \\
\sum_{\nu_{1} \, =\, 0}^{3} \sum_{{n_{2} \, =\, 1}}^{4} \left(
\sum_{\nu_{2} \, =\, 0}^{3} (\gamma^{\nu_{1}} \gamma^{\nu_{2}}
\gamma^{0})_{m_{2} n_{2}} \left( - i\frac{\partial}{\partial
x^{\nu_{2}}} \right) - (\gamma^{\nu_{1}} \gamma^{0} \beta
(\mu_{2}^{2}))_{m_{2} n_{2}} \right) \times \nonumber
\\ (f_{2})_{n_{2}} (K_{21;0}Q_{\nu \nu_{1}}^{0}
(\partial_{x}) + K_{21;1}Q_{\nu \nu_{1}}^{1} (\partial_{x}))
(x,x)^{n + 1}.
\end{eqnarray}
It is possible to neglect the vector potential (\ref{3.281}) in the
right - hand side of the equality  (\ref{3.27}) for the small
$\mu_{2}$. For $\mu_{2} = 0$ the vector potential (\ref{3.27}) is
simply equal to the vector potential (\ref{3.28}).

The equation (\ref{3.19}) is written for the functions
$(f_{1})_{n_{1}} (x_{1})$ equal to zero for the negative
$x_{1}^{0}$. Let us consider in the open upper light cone the
equation of the type (\ref{3.19}) for the arbitrary functions
$(\psi_{1})_{m_{1}} (x_{1})$ and the vector potential (\ref{3.28})
\begin{equation}
\label{3.29} \sum_{n_{1}\, =\, 1}^{4} \sum_{\nu \, =\, 0}^{3}
(\gamma^{\nu})_{m_{1} n_{1}} \left( - i\frac{\partial}{\partial
x_{1}^{\nu}} + B_{\nu}^{(21;0)} \right)(\psi_{1})_{n_{1}} (x_{1}) +
\sum_{n_{1}\, =\, 1}^{4} (\beta (\mu_{1}^{2}))_{m_{1} n_{1}}
(\psi_{1})_{n_{1}} (x_{1}) = 0.
\end{equation}
The equation (\ref{3.29}) is similar to the equation (\ref{1.2}).
Let the matrix $\alpha (\mu_{1})$ act on the equation (\ref{3.29}).
In view of the relations (\ref{2.15}) we have
\begin{equation}
\label{3.30} \sum_{n_{1}\, =\, 1}^{4} \sum_{\nu \, =\, 0}^{3}
(\gamma^{\nu})_{m_{1} n_{1}} \left( - i\frac{\partial}{\partial
x_{1}^{\nu}} + B_{\nu}^{(21;0)} \right)(\xi_{1})_{n_{1}} (x_{1}) +
\mu_{1} (\xi_{1})_{m_{1}} (x_{1}) = 0,
\end{equation}
\begin{equation}
\label{3.31} (\psi_{1})_{m_{1}} (x_{1}) =  \sum_{n_{1}\, =\, 1}^{4}
(\beta (\mu_{1}^{- 1}))_{m_{1} n_{1}} (\xi_{1})_{n_{1}} (x_{1}).
\end{equation}
Then the solution of the equation (\ref{3.29}) in the open upper
light cone is
\begin{equation}
\label{3.32} (\psi_{1})_{m_{1}} (x_{1}) =  \sum_{n_{1}\, =\, 1}^{4}
\sum_{k\, =\, 0}^{\infty} (k!)^{- 1} (- ix_{1}^{0})^{k} (\beta
(\mu_{1}^{- 1})C^{k})_{m_{1} n_{1}} (\xi_{1})_{n_{1}}
\end{equation}
where a vector$(\xi_{1})_{n_{1}}$ is independent of the variable
$x_{1}$ and the matrix
\begin{equation}
\label{3.33} C_{m_{1} n_{1}} = \sum_{\nu \, =\, 0}^{3}
B_{\nu}^{(21;0)} (\gamma^{0} \gamma^{\nu})_{m_{1} n_{1}} + \mu_{1}
(\gamma^{0})_{m_{1} n_{1}}
\end{equation}
is Hermitian if all components of the vector potential (\ref{3.28})
are real.

The vector potential (\ref{3.28}) and the mass $\mu_{1}$ define the
energy spectrum of the solution (\ref{3.32}). Choosing  the vectors
$(f_{2})_{n_{2}}$ it is possible to obtain a practically arbitrary
energy spectrum.

The vector potential (\ref{3.25}), $(f_{2})_{n_{2}} (+ 0,{\bf
x}_{2})$ $= (f_{2})_{n_{2}} \delta ({\bf x}_{2})$, is equal to zero
out of the closed upper light cone. The functions $(f_{1})_{n_{1}}
(x_{1})$ satisfy the free equation  (\ref{2.32}) out of the closed
upper light cone. If the initial wave function $(f_{1})_{n_{1}} (+
0,{\bf x}_{1})$ is equal to $(f_{1})_{n_{1}} \delta ({\bf x}_{1})$
(at the initial moment both particles are at the coordinates
origin), then in view of the relation (\ref{2.39}) the function
$(f_{1})_{n_{1}\dot{n}_{1}} (x_{1})$ is equal to zero out of the
closed upper light cone. We study the equation (\ref{3.19}) with the
singular vector potential (\ref{3.25}), $(f_{2})_{n_{2}} (+ 0,{\bf
x}_{2})$ $= (f_{2})_{n_{2}} \delta ({\bf x}_{2})$ and the singular
initial wave function $(f_{1})_{n_{1}} (+ 0,{\bf x}_{1})$ $=
(f_{1})_{n_{1}} \delta ({\bf x}_{1})$. The solution of the equation
(\ref{3.19}) is also singular. The solution singularities lie on the
upper light cone boundary.

\end{document}